\documentclass[sigplan,screen,pbalance]{acmart}

\usepackage{blindtext}
\usepackage{cleveref}
\usepackage{graphicx}
\usepackage{hyperref}
\usepackage{listings}
\usepackage{subfiles}
\usepackage{todonotes}
\usepackage{subcaption}
\usepackage{multirow}
\usepackage{mathpartir}
\usepackage{amsmath,gensymb}

\graphicspath{{images}}

\newcommand{\F}[1]{\text{\textsf{#1}}\xspace}

\newcommand{\benchmarks}{$61$\xspace}

\newcommand{\dedupA}{$2913$\xspace} 
\newcommand{\dedupB}{$4071$\xspace} 
\newcommand{\dedupC}{$102$\xspace} 
\newcommand{\dedupD}{$4$\xspace} 

\newcommand{\synthA}{$7215$\xspace} 
\newcommand{\synthB}{$4302$\xspace} 
\newcommand{\synthC}{$231$\xspace} 
\newcommand{\synthD}{$129$\xspace} 
\newcommand{\synthE}{$125$\xspace} 

\newcommand{\trivial}{$19$\xspace}
\newcommand{\useful}{$51$\xspace}
\newcommand{\definition}{$55$\xspace}
\newcommand{\nontrivial}{$106$\xspace}

\newcommand{\reductionA}{$96$\%\xspace}

\newcommand{\reductionB}{$44$\%\xspace}

\usepackage{booktabs}   
\usepackage{subcaption} 

\setcopyright{rightsretained}
\acmPrice{}
\acmDOI{10.1145/3520308.3534506}
\acmYear{2022}
\copyrightyear{2022}
\acmSubmissionID{pldiws22egraphsmain-p10-p}
\acmISBN{978-1-4503-9270-9/22/06}
\acmConference[EGRAPHS '22]{Proceedings of the 1st ACM SIGPLAN International Symposium on E-Graph Research, Applications, Practices, and Human-factors}{June 14, 2022}{San Diego, CA, USA}
\acmBooktitle{Proceedings of the 1st ACM SIGPLAN International Symposium on E-Graph Research, Applications, Practices, and Human-factors (EGRAPHS '22), June 14, 2022, San Diego, CA, USA}


\begin{document}

\title{Synthesizing Mathematical Identities with E-Graphs}         


\author{Ian Briggs}
\affiliation{
  \department{School of Computing}              
  \institution{University of Utah}            
  \city{Salt Lake City}
  \state{UT}
  \country{USA}
}
\email{ibriggs@cs.utah.edu}          

\author{Pavel Panchekha}
\affiliation{
  \department{School of Computing}              
  \institution{University of Utah}            
  \city{Salt Lake City}
  \state{UT}
  \country{USA}
}
\email{pavpan@cs.utah.edu}         

\begin{abstract}
Identities compactly describe properties of a mathematical expression
  and can be leveraged into faster and more accurate
  function implementations.
However, identities must currently be discovered manually,
  which requires a lot of expertise.
We propose a two-phase synthesis and deduplication pipeline
  that discovers these identities automatically.
In the synthesis step,
  a set of rewrite rules is composed, using an e-graph,
  to discover candidate identities.
However, most of these candidates are duplicates,
  which a secondary de-duplication step discards
  using integer linear programming and another e-graph.
Applied to a set of \benchmarks benchmarks,
  the synthesis phase generates \synthA candidate identities
  which the de-duplication phase then reduces down
  to \synthE core identities.
\end{abstract}


\maketitle

\section{Introduction}

Identities are a compact way of describing the properties
  of a mathematical expression.
For example,
  the $\sin(x)$ function is odd and periodic,
  which can be expressed via the identities
  $\sin(x) = -\sin(-x)$ and $\sin(x) = \sin(x + 2 \pi k)$.
Identifying the identities true of a particular expression
  allow one to write faster and more accurate
  implementations of that function.
We propose automatically synthesizing the identities
  necessary for range reduction and reconstruction of compound functions
  using e-graphs.

Specifically, this paper considers the task
  of synthesizing equalities $f(x) = s(f(t(x)))$
  from arbitrary mathematical expressions $f$
  in one variable $x$.
To do so, we use a set of rewrite rules,
  based on the Herbie floating-point synthesis tool~\cite{herbie},
  to generate equivalent forms of $f(x)$ in an e-graph,
  and extract expressions of the form $s(f(t(x)))$.
On a set of \benchmarks benchmarks,
  this generates \synthA identities.
The vast majority of extracted expressions are,
  however, duplicates of each other,
  and represent the same mathematical property.
We therefore de-duplicate these identities in a second egraph,
  which uses the same set of rewrite rules
  but, crucially, treats $f$ abstractly.
This means that identities are considered duplicates
  if they are equivalent for all possible functions $f$,
  that is, if they express the same property of $f$.
Deduplication significantly cuts down (by \reductionA)
  on the number of synthesized identities.
We further add a second de-duplication phase
  that considers compositions of identities
  (of the form $f(x) = s_1(s_2(f(t_2(t_1(x)))))$)
  and reduces the number of identities by a further \reductionB.
These synthesis and deduplication phases
  allow us to automatically synthesize
  a small yet descriptive set of identities
  for an arbitrary mathematical expression.

\section{Motivation}
\label{sec:motivation}

\begin{figure}
\includegraphics[width=\linewidth]{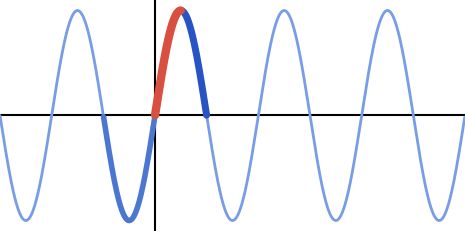}

\caption{
  A graph of $\sin(x)$,
    highlighting the portion over the domain $[0, \pi/2]$
    and indicating how various graphical symmetries
    allow one to recover the complete graph.
}
\label{fig:sineGraph}
\end{figure}

To implement a transcendental mathematical function in floating point, such as
  \F{sin}, \F{exp}, or \F{log}, an approximation must be used;
  some common techniques include
  polynomial approximation, table based interpolation, or combinations of these.
These approaches work particularly well
  when the input is drawn from a small domain,
  since polynomial-based and table-based approximations
  are typically less accurate when used over larger domains.
As such, implementations of mathematical functions
  require \textit{range reduction and reconstruction},%
\footnote{This technique is called range reduction and reconstruction
  despite modifying the domain, not the range, of the function in question.}
  which implement one function
  in terms of an auxiliary function over a smaller domain.
Range reduction and reconstruction uses the features of a function
  to derive an appropriate auxiliary function
  and map the desired function to it.
For example, consider the task of implementing $\sin(x)$.
It's relatively straight-forward to approximate $\sin(x)$
  over a narrow range like $[0, \pi/2]$ using a technique like
  Remez approximation~\cite{remez},
  but these techniques get less accurate as the domain grows larger,
  and are totally unworkable if, for example,
  the input $x$ could be any double-precision value.
However, as shown in \Cref{fig:sineGraph},
  the graph of $\sin(x)$ has many symmetries and self-similarities.
For example, $\sin(x)$ is horizontally symmetric,
  with the function looking identically to the left and right
  of the peak at $\pi/2$.
This means that to implement $\sin(x)$ over $[0, \pi]$,
  inputs between $\pi/2$ and $\pi$
  can be remapped to the input $[0, \pi/2]$.
In other words, an implementation of $\sin(x)$ over $[0, \pi]$
  can be reduced to an implementation over $[0, \pi/2]$.
Likewise, the left half of the graph
  is just a $180\degree$ rotation of right half.
Thus implementing $\sin(x)$ over $[-\pi, \pi]$
  can be reduced to implementing it over $[0, \pi]$,
  and negating the result for negative inputs.
Finally, $\sin(x)$ is periodic with period $2\pi$.
Thus, to implement $\sin(x)$ for arbitrary real numbers $x$,
  it's sufficient to take the input $x$ modulo $2\pi$,
  mapping the input into an arbitrary input range of width $2\pi$
  like, for example, $[-\pi, \pi]$.

Intuitively, these range reduction and reconstruction steps
  correspond to graphical symmetries of $\sin(x)$,
  but mathematically, these symmetries correspond
  to mathematical identities of $\sin(x)$.
Horizontal symmetry around $\pi/2$
  corresponds to the identity $\sin(x) = \sin(\pi - x)$;
  the $180\degree$ rotation corresponds to $\sin(x) = -\sin(-x)$;
  and periodicity corresponds to the fact
  that $\sin(x) = \sin(x - 2 \pi n)$ for any integer $n$.
Each of these identities correspond to a step in our $\sin(x)$ implementation:
  horizontal symmetry means subtracting inputs in $[\pi/2, \pi]$ from $\pi$;
  the $180\degree$ symmetry means negative inputs
  (and storing a flag to negate the results before returning);
  and periodicity requires computing the input $x$ modulo $2\pi$.
Other $\sin(x)$ identities, like the double-angle formula,
  could also be used for range reduction and reconstruction.

More generally, implementions of mathematical functions
  typically leverage identities of those functions
  to operate over a larger set of inputs or to improve accuracy and speed.
In the general case, these identities take the form
  $f(x) = s(f(t(x)))$
  where $t$ corresponds to the range reduction function
  and $s$ corresponds to the reconstruction function.
For example, for $\sin(x)$, the $180\degree$ rotation identity
  has $t(x) = -x$ and $s(y) = -y$,
  while the periodicity identity has
  $t(x) = x + 2\pi$ and $s(y) = y$;
  \ref{tbl:reductionTable} lists $s$ and $t$ for $\sin(x)$'s other identities.

\begin{table}
\caption{
Four identities true for $\sin(x)$,
  and their representation in the form
  $\sin(x) = s(\sin(t(x)))$.
For two of the identities $s(y) = y$,
  in which case no reconstruction step is needed.
}

\begin{tabular}{lllll}
Inputs               & Outputs        & $sin(x)=$          & $s(y)=$  & $t(x)=$ \\ \hline
$[-\infty, \infty]$  & $[0, \infty]$  & $-\sin(-x)$        & $-y$     & $-x$ \\
$[0, \infty]$        & $[0, 2\pi]$    & $\sin(x - n2\pi)$  & $y$      & $x - 2\pi\lfloor \frac{x}{2 \pi} \rfloor$ \\
$[0, 2\pi]$          & $[0, \pi]$     & $-\sin(x-\pi)$     & $-y$     & $x-\pi$ \\
$[0, \pi]$           & $[0, \pi/2]$   & $sin(\pi-x)$       & $y$      & $\pi-x$ \\
\end{tabular}
\label{tbl:reductionTable}
\end{table}

Unfortunately, today, range reduction and reconstruction steps
  are coded manually by a mathematical expert,
  even though the actual polynomial-based or table-based core
  can be derived with automated tools~\cite{sollya,rlibm}.
For elemental functions such as \F{sin}, \F{log}, or \F{exp},
  these identities are relatively common knowledge;
  but for compound mathematical functions
  like  $log(x+1) - log(x)$,
  $(1 - (tan(x) \cdot tan(x))) / (1 + (tan(x) \cdot tan(x)))$,
  or $(1 - cos(x)) / sin(x)$,
  the useful identities might not be so obvious.
This means that high-quality implementations of these compound functions
  are still the domain of experts with deep experience and knowledge.

\section{Synthesizing}

Our overall approach is to construct an e-graph
  containing the compound function $f(x)$
  and use the axioms of basic functions as rewrite rules
  to discover identities about $f(x)$ like $f(x) = -f(x + k)$.
Then, all e-nodes in the e-class of $f(x)$ are extracted
  and any with the form $s(f(t(x)))$ are taken as candidate identities.

\subsection{Grammar and Rewrite Rules}

To do any of this we need to start with a firm e-graph world to stand on,
  meanting a grammar and accompanying rewrite rules
  so that combinations of rewrites can discover meaningful identities.
Our grammar starts with standard mathematical operations
  such as addition, subtraction, multiplication, division,
  as well as common mathematical functions such as
  square root, trigonometric functions, exponential functions, and logarithmic functions.
The compound functions whose identities we seek
  will be defined in terms of these common operations.
We assume these operations ultimately apply to real-valued constants
  and a single real-number argument $x$.
Then, mathematical identities
  such as $x + y = y + x$ or $\cos(x) = \cos(-x)$
  are encoded as rewrite rules.

Initially, we derived our rule set from the Herbie expression simplifier~\cite{herbie},
  but we quickly realized that that rule set was unusable for our purposes
  because it contains many unsound rules.
For example, consider the rule $a \cdot (1/a) \leadsto 1$,
  called \texttt{rgt-mult-inverse} in Herbie.
While innocous-looking, it rewrites the expression $0 \cdot (1 / 0)$
  to $1$, while the equally-innocuous rule \texttt{mul0-lft},
  which states that $0 \cdot a \leadsto 0$,
  rewrites $0 \cdot (1 / 0)$ to $0$.
In an e-graph, using both of these rules simultaneously
  allows one to prove that $0 = 1$,
  which is both untrue and causes significant problems for extraction.
In Herbie, this is not a significant issue because the rules are run
  for few iterations, and derivations like this are rarely hit.
Moreover, Herbie saves the state of its e-graph
  and rewinds to an earlier iteration
  if two obviously-different things are proven equal.
However, for our task, a different approach is needed,
  since runs with many iterations are necessary
  to discover valuable identities about compound functions.

We therefore need to choose a subset of rules
  where derivations like the above are impossible.
In the example above, the issue is clearly the input expression,
  $0 \cdot (1 / 0)$, which is undefined everywhere.
We thus need to ensure that none of our rewrite rules
  can ever create expressions of this form;
  Table~\ref{tbl:invalidDomains} lists functions in our grammar
  and the domains that they are not defined over.
For example, the rewrite rule $a / b \leadsto 1 / (b / a)$ must be dropped,
  since applying it to the safe, well-defined expression $0 / 1$
  constructs the ill-defined term $1 / (1 / 0)$.
Mathematically, the issue is that the left hand side of this rule
  is defined as long as $b$ is nonzero,
  while the right hand side additionally requires that $a$ is non-zero
  in order to be defined.
More generally, for each of our rewrite rules,
  we need to ensure that the right hand side is defined at all points
  that the left hand side is defined at;
  if this is true, then inductively every expression in the e-graph
  is defined at all points that the original compound function is
  defined at,
  and wholely-undefined expressions are never generated.
This typically requires some kind of conditional reasoning.
For example, in the rule $a/(b\cdot c) \leadsto (a/b)/c$,
  we can assume that the left hand side is defined
  and therefore that $b\cdot c$ does not equal zero (since we divide by it),
  while on the right hand side we need to prove
  that $b$ is not equal to zero (so that we can divide by it)
  and that $c$ is not equal to zero (so that we can also divide by it).
Luckily, $b\cdot c$ is nonzero if and only if both $b$ and $c$ are nonzero,
  so this rule is safe to apply.
We applied similar reasoning to each of the rules in the Herbie rule set,
  filtering it down to a set of safe rules that cannot lead
  to the unsound equivalences described above.

\begin{table}
\caption{Operations which can be undefined.}
\label{tbl:invalidDomains}

\begin{tabular}{lllll}
Operation            & Invalid Domain \\ \hline
$a/b$                & $b = 0$ \\
$\F{acos}(x)$        & $x < -1 \lor 1 < x$ \\
$\F{acosh}(x)$       & $x < 1$ \\
$\F{asin}(x)$        & $x < -1 \lor 1 < x$ \\
$\F{log}(x)$         & $x \le 0$ \\
$\F{log1p}^{-1}(x)$  & $x \le -1$ \\
$\F{sqrt}(x)$        & $x < 0$ \\
$\F{atan2}(y, x)$    & $x = 0$ \\
\end{tabular}
\end{table}

\subsection{E-Graphs for Identities}

With the grammar and rewrite rules set,
  e-graphs provide a straightforward way to derive a whole lot of
  equivalent formulations of the compound function.
For example, if the compound function is $f(x) = \tan(x) - \sin(x)$,
  our rewrite rules allow us to prove it equal to
  $-(\tan(-x) - \sin(-x))$,
  or in other words that $f(x) = -f(-x)$.
Notice that we are interested only in formulations of $f(x)$
  that themselves are phrased in terms of calls to $f$,
  so standard e-graph extraction can't be used.
Instead, we want to artificially lower the cost of calls to $f$,
  so that equivalent formulations that call $f$ are preferred.
This is tricky to do, since $f$ is given by a compound expression,
  while extraction looks at e-nodes one at a time,
  and so can't know if a call to $f$ is even being considered.
To get around this issue, we introduce a new operator to our grammar,
  $\F{thefunc}(x)$,
  plus new rewrite rules representing the equality $\F{thefunc}(x) = f(x)$.
Note that this equality is the \textit{only} equality constraining $\F{thefunc}$.
This way, we not only prove $\tan(x) - \sin(x)$ equal to $-(\tan(-x) - \sin(-x))$,
  but also $\F{thefunc}(x) = -\F{thefunc}(-x)$.
Extraction can then preferentially select expressions containing \F{thefunc}
  by setting the cost of the \F{thefunc} operator to zero during extraction.

Traditionally, e-graph extraction is used to select
  the single simplest expression represented in the e-graph
  and equal to a given starting point.
However, in our case, we want to extract multiple expressions
  representing a diverse set of different identities that can be combined
  into a useful range reduction and reconstruction algorithm.
This requires a twist on the traditional extraction algorithm.
In a traditional e-graph extraction,
  a simplest (lowest-cost) form is computed for every e-class in the e-graph,
  and the simplest form of the initial e-class is returned.
This results in the simplest form of the initial expression.
Since we instead want many different formulations
  of the initial expression $\F{thefunc}(x)$,
  we are looking to extract multiple expressions from a single e-class;
  but we also don't want our extracted expressions to contain
  unnecessary junk like $\F{thefunc}(x + 0 \cdot (\ldots))$.
To balance the goals of diversity and simplicity,
  we made a custom extractor that returns
  standard extractions of \textit{all enodes} in the initial e-class.
Each of those e-nodes represents a different formulation of the initial expression,
  since duplicate e-nodes are merged in an e-graph;
  however, the standard extraction of each of those e-nodes
  uses simplest form for each \textit{argument} of the e-node,
  meaning that the extracted expressions are all still relatively simple.

The combination of zero cost for $\F{thefunc}$ nodes
  and standard extractions of each e-node in the initial e-class
  allows us to extract many different formulations of $\F{thefunc}(x)$
  that preferentially contain calls to $\F{thefunc}$,
  and many of the extracted expressions represent identities of $f(x)$.
However, these extracted expressions also contain
  many duplicates and redundant identities.

\section{Deduplicating}

In range reduction and reconstruction,
  duplicate or redundant identities are never useful,
  so should be automatically removed.
For instance, the identity $f(x) = 0 + f(x)$
  does not help reduce the range of $x$,
  yet is represented by a different e-node (a plus node)
  than the initial expression $\F{thefunc}(x)$.
However, some identities that do not change the range of $x$,
  such as $f(x) = \F{fabs}(f(x))$,
  do present information useful during range reduction,
  in this case that $f(x)$ is uniformly positive.
What determines if an extracted identity is useful?
Our insight is that identities like $f(x) = 0 + f(x)$ are useless
  precisely because they are true of every possible function $f(x)$,
  while an identity like $\F{fabs}(f(x))$
  is true about only some functions $f(x)$
  and therefore provides non-trivial information about $f$.
This insight allows us to use an e-graph for deduplication.

Specifically, we create a new, secondary e-graph
  containing all of the expressions extracted from the first e-graph,
  expressed in terms of $\F{thefunc}(x)$ and initially unequal,
  including the expression $\F{thefunc}(x)$.
We then run the same set of rewrite rules
  except for the rules representing the equality $\F{thefunc}(x) = f(x)$.
By withholding this equality,
  we effectively ask the e-graph to prove
  which of our extracted expressions are the same
  for all possible meanings of $\F{thefunc}$.
Any extracted expressions proved equal in this e-graph
  are true for all possible functions $f$ and are therefore duplicates.
Thus, to eliminate duplicates,
  we merely re-extract all of the expressions from this new e-graph,
  which will yield the same extraction
  for expressions that are equal for all possible $f$,
  and throw away the duplicates.

A similar approach can be used to identify not only duplicate rules
  but also rules that are compositions of other rules.
For example, for $f(x) = \tan(x) - \sin(x)$,
  both $I_1 = [f(x) = f(x + 2\pi)]$ and $I_2 = [f(x) = f(x + 4\pi)]$
  are true identities,
  and moreover these two identities are not equivalent
  for arbitrary functions $f(x)$.
However, the first identity, applied twice, results in the second identity,
  and in fact the first identity can be repeated any number of times
  to result in $f(x) = f(x + 2 \pi n)$.
This means that any use of the second identity during range reduction
  can always be better accomplished using the first identity,
  and so the second identity ought to be filtered out.
We do so using e-graphs and integer linear programming.

Like for deduplicating, we create a new e-graph
  containing all the extracted expressions,
  expressed in terms of $\F{thefunc}(x)$ and initially unequal,
  as well as all pairwise compositions of these expressions,
  where composing one expression with another
  means substituting the second expression
  into all uses of $\F{thefunc}$ in the first expression.
Applying rewrite rules in this e-graph
  allows us to prove identities like $I_1 \circ I_1 = I_2$,
  and re-extracting all composed expressions
  allows us to identify which composed identities
  are equal to which non-composed identities.
In this case, we say the identity $I_2$ is covered
  by the set $\{I_1\}$ of identities,
  in the sense that it is equal to a composition
  of identities from that set.
We now aim to select a minimal set of core identities
  that cover all the discovered identities.

This is a kind of set-cover problem, which we encode
  as integer linear programming like so.
For each identity we introduce two variables:
  $I_n$, which indicates whether the identity is part of the core set,
  and $cI_n$, indicates whether it is covered by identities in the core set.
These variables are constrained by the equivalences discovered by the e-graph:
  if $I_i \circ I_j = I_k$, then $cI_i \land cI_j \implies cI_k$.
All identities in the core set are covered as well: $I_i \implies cI_i$.
To put this in more concrete terms,
  for $f(x) = \tan(x) - \sin(x)$,
  the constraint for $I_2$ reads
  $cI_2 = I_2 \lor (cI_1 \land cI_1)$.
The integer linear program is then asked to minimize
  the sum of the $I_n$s, that is, minimize the number of core identities.

However, additional constraints are needed to prevent ``cyclic'' reasoning
  in the case of equalities like $I_1 \circ I_1 = I_1$,
  where the constraint $cI_1 = I_1 \lor (cI_1 \land cI_1)$
  is satisfiable with $I_1 = \bot$ and $cI_1 = \top$.
To avoid this, we enforce a kind of provenance
  where each covered identity must be covered
  by a finite sequence of compositions from the core set.
Each identity now gets a positive age variable, $aI_n$,
  which defines when the identity was covered.
If an identity is in the core set (meaning $I_n$ is true)
  then its age $aI_n = 1$.
Otherwise, its age is the sum of the ages of the covered identities
  that compose to it;
  for the $\tan(x) - \sin(x)$ case, this means
  the constraint reads:
\[
cI_2 = (I_2 \land aI_2 = 1) \lor (cI_1 \land cI_1 \land aI_2 = aI_1 + aI_1)
\]
The age variables eliminate the possibility of cycles;
  in the case of $I_1 \circ I_1 = I_1$,
  the constraint reads:
\[
cI_1 = (I_1 \land aI_1 = 1) \lor (cI_1 \land cI_1 \land aI_1 = aI_1 + aI_1)
\]
The equation $aI_1 = aI_1 + aI_1$ is unsatisfiable,
  since all ages are positive,
  eliminating cycles and
  meaning that setting $cI_1$ requires setting $I_1$.
The addition of ages transforms this from a SAT problem into an integer linear
  programming problem, but luckily one that is typically small and easy to solve.
Using this encoding stops self supporting logic from forming, since the age of
  any covered identity can't be less than the source of that cover.

The solution to this integer linear programming problem
  identifies a set of core identities of $f(x)$
  that are not duplicates
  and that can be composed to derive
  any other identity of $f(x)$ provable from our set of rewrite rules.
These core identities can then be presented to the user or,
  eventually, integrated into an end-to-end mathematical function
  synthesis tool.

\section{Results}


%

\begin{figure}
\includegraphics[width=\linewidth]{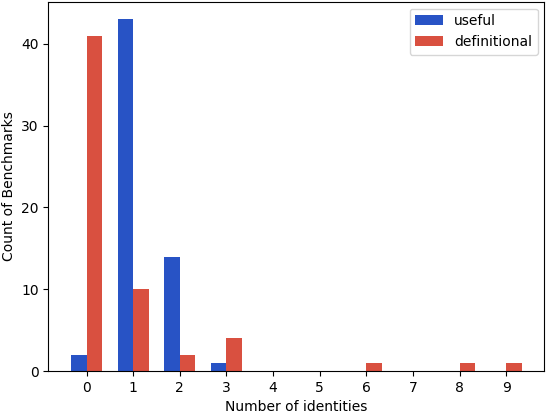}

\caption{
A split histogram of number of identities found,
  with blue bars counting the number of benchmarks
  with that many useful identities
  and red bars counting the number of benchmarks
  with that many definitional (useless) identities.
Starting at the left there are $2$ benchmarks for which we found no good
  identities and $41$ benchmarks that had no bad identities.
Note that most benchmarks have at least one useful identity,
  while relatively few benchmarks have definitional identities
  (though some have very many of them).
}
\label{fig:histogram}
\end{figure}

We implemented this approach on top of the egg e-graph library~\cite{egg}
  via the \F{snake\_egg} Python package
  and applied it to \benchmarks mathematical expressions
  from the FPBench~\cite{fpbench} and Herbie~\cite{herbie} benchmark suites,
  plus additional expressions defining variant trigonometric functions
  like \F{versin}, \F{havercosin}, and similar.
In total, the first synthesis step synthesized \synthA expressions across all benchmarks.
Of these \dedupA contained no form of their target function, leaving
  \synthB candidate identities.
Deduplication then removes \dedupB identities
  that are true for all possible functions $f$,
  leaving \synthB identities that provide useful information about the function.
Note that the vast majority of candidate identities are duplicates,
  for the simple reason that these identities can be found
  for any possible input program,
  while the non-trivial and non-duplicate identities
  require actual reasoning about the function at hand.
Deduplicating compound functions using integer linear programming
  removed \dedupC, leaving a penultimate count of \synthC.
Of these, \trivial are the trivial identity of $thefunc(x) = thefunc(x)$,
  which is true of all benchmarks but is not always present
  because it can sometimes be represented as a compound of two other identities:
  for example, if $I = [f(x) = -f(-x)]$,
  then $I \circ I = [f(x) = --f(--x)]$,
  which simplifies to the trivial identity $[f(x) = f(x)]$.
In this case, the trivial identity won't be in the minimal core set.
This means that of the \synthD deduplicated identities, \nontrivial are non-trivial,
  shown in \Cref{fig:histogram}.
Of these, a healthy \useful of them, just under half,
  look, upon manual examination,
  to correspond to useful range reductions.
This means that our approach is effective
  at automatically synthesizing identities
  that are useful for range reduction and reconstruction of compound functions.

The other \definition non-trivial identities are, however, less useful,
  in that they just encode the definition of $f(x)$ in a complicated form.
For example, consider $f(x) = 1 + \cos(x)$,
  for which our tool generates the identity
  $(1 - thefunc(x)) - (-thefunc(x) - cos(x))$.
Distributing the subtraction, we get $1 - thefunc(x) + thefunc(x) + cos(x)$,
  and the two instances of $thefunc(x)$ cancel
  to leave just $1+cos(x)$, the definition of $f(x)$.
This is not a useful identity of $f(x)$ and does not suggest
  a possible range reduction and reconstruction approach.
This issue is, in a sense, the dual of duplication and redundancy:
  where duplicate identities are those that are equivalent
  even not knowing the definition of $f(x)$,
  these definitional identities are those that are \textit{only} true
  for the single, fixed $f(x)$.
In this example, the definition identity can be detected by noticing
  that it is equivalent, for arbitrary $f(x)$,
  to an expression that doesn't use $f(x)$ at all,
  meaning that calling $f(x)$ is unnecessary and that this identity
  isn't helpful for range reduction.
In the egg library, an e-graph analysis can be used
  to determine whether $f(x)$ is equivalent
  to an expression that does not call $f$.

However, not all definitional identities have this form;
  for example, consider $f(x) = \sin(x)/2$,
  where our approach produces the identity $f(x) = \sin(x) - f(x)$.
Here, the right hand side can't be simplified further,
  but the \textit{equality as a whole} can be rewritten to
  $2 f(x) = \sin(x)$
  and then to $f(x) = \sin(x) / 2$,
  where the right-hand side again does not call $f(x)$.
We do not yet have a comprehensive approach
  to detecting and eliminating these kinds of definitional identities.
Our current approach attempts to derive the definition of $f(x)$
  from each identity, keeping only the identities where this is impossible.
To do so, we might, for example,
  take the right-hand-side $s(f(t(x)))$ of the identity
  and adds the equality $f(x) - s(f(t(x))) = 0$ to an e-graph.
If $f(x)$ can then be proven equal to an expression
  without any calls to $f$, the identity is definitional and can be discarded.
In our experiments, this approach removes just \dedupD definitional identities,
  suggesting that it is not a particularly effective method
  for dealing with definitional identities.
It's possible that variantions on this approach
  (like adding the equality $f(x) / s(f(t(x))) = 1$),
  will be more effective, but that's not clear at the moment.
Despite the presence of many definitional identities,
  our approach is clearly surfacing many true and useful identities
  and provides useful assistance to a programmer
  implementing a mathematical function.

\section{Discussion \& Related Work}

Our approach is similar to existing work on
  rewrite rule synthesis, algebraic rewriting,
  and library function implementation.

Ruler~\cite{ruler} is a synthesis tool
  that generates rewrite rules over an arbitrary domain $D$.
For example, given an interpreter for
  arithmetic expressions on rational numbers,
  Ruler can automatically synthesize facts
  like commutativity, associativity, and the difference of squares.
Like Ruler, our approach generates mathematical equalities
  over a domain, though unlike in Ruler,
  our domain includes a function symbol $thefunc$.
Moreover, while Ruler-generated rewrite rules
  can be arbitrary expressions from its grammar,
  our approach generates rules strictly of the form
  $f(x) = s(f(t(x)))$.
Because Ruler rules can have arbitrary
  left- and right-hand sides,
  in its synthesis step it must consider
  all pairs of e-nodes in the e-graph,
  while in our more restricted setting
  we can use ordinary e-graph extraction.
Ruler attempts to minimize the set of rules it synthesizes,
  but its approach is different from ours.
Ruler uses an approach similar to delta-debugging,
  where subsets of synthesized rules are tested
  to check whether they can compose to form
  the full set of synthesized rules.
One can see this as one heuristic method
  to approximately solve
  the ILP problem our approach uses for minimization.
We expect fewer rewrites to be available in our domain,
  and desire more minimization,
  so the slower but more optimal ILP solution
  is more appropriate.
Finally,
  Ruler requires a complete interpreter and verifier for its domain,
  which makes it difficult (or perhaps impossible?)
  to apply to domains like exponential or trigonometric functions.
Our approach is instead based on basic mathematical axioms,
  so has no trouble with exponential or trigonometric functions.

In algebraic rewriting,
  identities are used to transform one mathematical expression
  into another.
For example, the Herbie floating-point synthesis tool~\cite{herbie}
  rewrites mathematical expressions using an e-graph
  in an attempt to find an equivalent expression
  with less floating-point error.
In Herbie~1.5, a feature was added
  that allows Herbie to leverage symmetric expressions.
For example, in the expression $f(a, b) = \sqrt{a^2 + b^2}$,
  the variables $a$ and $b$ are symmetric,
  meaning that swapping $a$ and $b$ results in the same value.
In such a case, Herbie automatically inserts a sorting step,
  which sometimes allows for more accurate results.
These symmetries could be seen
  as an instance of our framework,
  with the identity reading $f(a, b) = f(b, a)$
  meaning $s(y) = y$ and $t(a, b) = (b, a)$.

Herbie's approach was an inspiration for this work,
  and we suspect that a generalization of the approach in this paper
  could discover symmetric expressions.
However, our current implementation
  does not handle multi-variable expressions;
  the success of symmetric expressions in Herbie
  suggests that extending our implementation
  would be valuable.
Such an extensions could potentially allow Herbie
  to handle a larger class of expressions,
  including anti-symmetric expressions
  $f(a, b) = -f(b, a)$
  or expressions that are both symmetric
  and have other identities;
  for example, $\sqrt{a^2 + b^2}$ is not only symmetric
  but also even in both $a$ and $b$.

Mathematical library implementation tools such as MetaLibm~\cite{metalibm},
  Flopoco~\cite{flopoco}, and RLibm~\cite{rlibm}
  provide utilities that assist experts in writing
  implementations of mathematical functions like $\sin(x)$.
However, all three of these tools
  require the expert to identify and leverage
  mathematical identities such as
  periodicity, evenness, or oddness.
Our approach potentially paves the way
  for a fully automated mathematical function implementation synthesis,
  where identities are automatically synthesized,
  deduplicated, and then used to generate
  faster and more accurate function implementations.
We plan to pursue this avenue
  as the next step for our implementation,
  combining it with established techniques~\cite{remez,sollya,rlibm}
  for generating polynomial- or table-based
  implementations of functions over a narrow range.

\bibliography{bibfile}


\begin{thebibliography}{9}


\ifx \showCODEN    \undefined \def \showCODEN     #1{\unskip}     \fi
\ifx \showDOI      \undefined \def \showDOI       #1{#1}\fi
\ifx \showISBNx    \undefined \def \showISBNx     #1{\unskip}     \fi
\ifx \showISBNxiii \undefined \def \showISBNxiii  #1{\unskip}     \fi
\ifx \showISSN     \undefined \def \showISSN      #1{\unskip}     \fi
\ifx \showLCCN     \undefined \def \showLCCN      #1{\unskip}     \fi
\ifx \shownote     \undefined \def \shownote      #1{#1}          \fi
\ifx \showarticletitle \undefined \def \showarticletitle #1{#1}   \fi
\ifx \showURL      \undefined \def \showURL       {\relax}        \fi
\providecommand\bibfield[2]{#2}
\providecommand\bibinfo[2]{#2}
\providecommand\natexlab[1]{#1}
\providecommand\showeprint[2][]{arXiv:#2}

\bibitem[Damouche et~al\mbox{.}(2016)]%
        {fpbench}
\bibfield{author}{\bibinfo{person}{Nasrine Damouche}, \bibinfo{person}{Matthieu
  Martel}, \bibinfo{person}{Pavel Panchekha}, \bibinfo{person}{Jason Qiu},
  \bibinfo{person}{Alex Sanchez-Stern}, {and} \bibinfo{person}{Zachary
  Tatlock}.} \bibinfo{year}{2016}\natexlab{}.
\newblock \showarticletitle{Toward a Standard Benchmark Format and Suite for
  Floating-Point Analysis}.
\newblock  (\bibinfo{year}{2016}).
\newblock


\bibitem[de~Dinechin and Pasca(2011)]%
        {flopoco}
\bibfield{author}{\bibinfo{person}{Florent de Dinechin} {and}
  \bibinfo{person}{Bogdan Pasca}.} \bibinfo{year}{2011}\natexlab{}.
\newblock \showarticletitle{Designing Custom Arithmetic Data Paths with
  {FloPoCo}}.
\newblock \bibinfo{journal}{\emph{{IEEE} Design \& Test of Computers}}
  \bibinfo{volume}{28}, \bibinfo{number}{4} (\bibinfo{date}{July}
  \bibinfo{year}{2011}), \bibinfo{pages}{18--27}.
\newblock


\bibitem[Kupriianova and Lauter(2014)]%
        {metalibm}
\bibfield{author}{\bibinfo{person}{Olga Kupriianova} {and}
  \bibinfo{person}{Christoph Lauter}.} \bibinfo{year}{2014}\natexlab{}.
\newblock \showarticletitle{Metalibm: {A} {Mathematical} {Functions} {Code}
  {Generator}}. In \bibinfo{booktitle}{\emph{Mathematical {Software} – {ICMS}
  2014}} \emph{(\bibinfo{series}{Lecture {Notes} in {Computer} {Science}})},
  \bibfield{editor}{\bibinfo{person}{Hoon Hong} {and} \bibinfo{person}{Chee
  Yap}} (Eds.). \bibinfo{publisher}{Springer}, \bibinfo{address}{Berlin,
  Heidelberg}, \bibinfo{pages}{713--717}.
\newblock
\showISBNx{978-3-662-44199-2}
\urldef\tempurl%
\url{https://doi.org/10.1007/978-3-662-44199-2_106}
\showDOI{\tempurl}


\bibitem[Lim and Nagarakatte(2021)]%
        {rlibm}
\bibfield{author}{\bibinfo{person}{Jay~P. Lim} {and} \bibinfo{person}{Santosh
  Nagarakatte}.} \bibinfo{year}{2021}\natexlab{}.
\newblock \showarticletitle{{RLIBM-ALL:} {A} Novel Polynomial Approximation
  Method to Produce Correctly Rounded Results for Multiple Representations and
  Rounding Modes}.
\newblock \bibinfo{journal}{\emph{CoRR}}  \bibinfo{volume}{abs/2108.06756}
  (\bibinfo{year}{2021}).
\newblock
\showeprint[arXiv]{2108.06756}
\urldef\tempurl%
\url{https://arxiv.org/abs/2108.06756}
\showURL{%
\tempurl}


\bibitem[Nandi et~al\mbox{.}(2021)]%
        {ruler}
\bibfield{author}{\bibinfo{person}{Chandrakana Nandi}, \bibinfo{person}{Max
  Willsey}, \bibinfo{person}{Amy Zhu}, \bibinfo{person}{Yisu~Remy Wang},
  \bibinfo{person}{Brett Saiki}, \bibinfo{person}{Adam Anderson},
  \bibinfo{person}{Adriana Schulz}, \bibinfo{person}{Dan Grossman}, {and}
  \bibinfo{person}{Zachary Tatlock}.} \bibinfo{year}{2021}\natexlab{}.
\newblock \showarticletitle{Rewrite Rule Inference Using Equality Saturation}.
\newblock \bibinfo{journal}{\emph{Proc. ACM Program. Lang.}}
  \bibinfo{volume}{5}, \bibinfo{number}{OOPSLA}, Article
  \bibinfo{articleno}{119} (\bibinfo{date}{oct} \bibinfo{year}{2021}),
  \bibinfo{numpages}{28}~pages.
\newblock
\urldef\tempurl%
\url{https://doi.org/10.1145/3485496}
\showDOI{\tempurl}


\bibitem[Pachón and Trefethen(2009)]%
        {remez}
\bibfield{author}{\bibinfo{person}{Ricardo Pachón} {and}
  \bibinfo{person}{Lloyd~N. Trefethen}.} \bibinfo{year}{2009}\natexlab{}.
\newblock \showarticletitle{Barycentric-{Remez} algorithms for best polynomial
  approximation in the chebfun system}.
\newblock \bibinfo{journal}{\emph{BIT Numerical Mathematics}}
  \bibinfo{volume}{49}, \bibinfo{number}{4} (\bibinfo{date}{Oct.}
  \bibinfo{year}{2009}), \bibinfo{pages}{721}.
\newblock
\showISSN{1572-9125}
\urldef\tempurl%
\url{https://doi.org/10.1007/s10543-009-0240-1}
\showDOI{\tempurl}


\bibitem[Panchekha et~al\mbox{.}(2015)]%
        {herbie}
\bibfield{author}{\bibinfo{person}{Pavel Panchekha}, \bibinfo{person}{Alex
  Sanchez-Stern}, \bibinfo{person}{James~R. Wilcox}, {and}
  \bibinfo{person}{Zachary Tatlock}.} \bibinfo{year}{2015}\natexlab{}.
\newblock \showarticletitle{Automatically improving accuracy for floating point
  expressions}. In \bibinfo{booktitle}{\emph{Proceedings of the 36th {ACM}
  {SIGPLAN} {Conference} on {Programming} {Language} {Design} and
  {Implementation}}} \emph{(\bibinfo{series}{{PLDI} '15})}.
  \bibinfo{publisher}{Association for Computing Machinery},
  \bibinfo{address}{New York, NY, USA}, \bibinfo{pages}{1--11}.
\newblock
\showISBNx{978-1-4503-3468-6}
\urldef\tempurl%
\url{https://doi.org/10.1145/2737924.2737959}
\showDOI{\tempurl}


\bibitem[{S. Chevillard} et~al\mbox{.}(2010)]%
        {sollya}
\bibfield{author}{\bibinfo{person}{{S. Chevillard}}, \bibinfo{person}{{M.
  Joldeş}}, {and} \bibinfo{person}{{C. Lauter}}.}
  \bibinfo{year}{2010}\natexlab{}.
\newblock \showarticletitle{Sollya: {An} {Environment} for the {Development} of
  {Numerical} {Codes}}. In \bibinfo{booktitle}{\emph{Mathematical {Software} -
  {ICMS} 2010}} \emph{(\bibinfo{series}{Lecture {Notes} in {Computer}
  {Science}}, Vol.~\bibinfo{volume}{6327})},
  \bibfield{editor}{\bibinfo{person}{{K. Fukuda}}, \bibinfo{person}{{J. van der
  Hoeven}}, \bibinfo{person}{{M. Joswig}}, {and} \bibinfo{person}{{N.
  Takayama}}} (Eds.). \bibinfo{publisher}{Springer},
  \bibinfo{address}{Heidelberg, Germany}, \bibinfo{pages}{28--31}.
\newblock


\bibitem[Willsey et~al\mbox{.}(2021)]%
        {egg}
\bibfield{author}{\bibinfo{person}{Max Willsey}, \bibinfo{person}{Chandrakana
  Nandi}, \bibinfo{person}{Yisu~Remy Wang}, \bibinfo{person}{Oliver Flatt},
  \bibinfo{person}{Zachary Tatlock}, {and} \bibinfo{person}{Pavel Panchekha}.}
  \bibinfo{year}{2021}\natexlab{}.
\newblock \showarticletitle{Egg: Fast and Extensible Equality Saturation}.
\newblock \bibinfo{journal}{\emph{Proc. ACM Program. Lang.}}
  \bibinfo{volume}{5}, \bibinfo{number}{POPL}, Article \bibinfo{articleno}{23}
  (\bibinfo{date}{jan} \bibinfo{year}{2021}), \bibinfo{numpages}{29}~pages.
\newblock
\urldef\tempurl%
\url{https://doi.org/10.1145/3434304}
\showDOI{\tempurl}


\end{thebibliography}

\end{document}